\documentclass[conference,10pt]{IEEEtran}
\IEEEoverridecommandlockouts
\usepackage{cite}
\usepackage{amsmath,amssymb,amsfonts}
\usepackage[ruled,linesnumbered]{algorithm2e}
\usepackage{graphicx}
\usepackage{dblfloatfix}
\usepackage{multirow}
\usepackage{comment}
\usepackage{textcomp}
\usepackage{xcolor}
\usepackage[a4paper, total={184mm,241.5mm}]{geometry}
\def\BibTeX{{\rm B\kern-.05em{\sc i\kern-.025em b}\kern-.08em
    T\kern-.1667em\lower.7ex\hbox{E}\kern-.125emX}}

\author{\IEEEauthorblockN{Jingxiao Ma, Sherief~Reda,~\IEEEmembership{Senior Member,~IEEE}}
\IEEEauthorblockA{\textit{School of Engineering, Brown University, Providence RI 02912} \\
}
}

\ifCLASSOPTIONcompsoc
    \usepackage[caption=false, font=normalsize, labelfont=sf, textfont=sf]{subfig}
\else
\usepackage[caption=false, font=footnotesize]{subfig}
\fi

\begin{document}

\title{RUCA: RUntime Configurable Approximate Circuits with Self-Correcting Capability}

\maketitle

\begin{abstract}
Approximate computing is an emerging computing paradigm that offers improved power consumption by relaxing the requirement for full accuracy. Since real-world applications may have different requirements for design accuracy, one trend of approximate computing is to design runtime quality-configurable circuits, which are able to operate under different accuracy modes with different power consumption. In this paper, we present a novel framework RUCA which aims to approximate an arbitrary input circuit in a runtime configurable fashion. By factorizing and decomposing the truth table, our approach aims to approximate and separate the input circuit into multiple configuration blocks which support different accuracy levels, including a corrector circuit to restore full accuracy. By activating different blocks, the approximate circuit is able to operate at different accuracy-power configurations.
To improve the scalability of our algorithm, we also provide a design space exploration scheme with circuit partitioning to navigate the search space of possible approximations of subcircuits during design time.  
We thoroughly evaluate our methodology on a set of benchmarks and compare against another quality-configurable approach, showcasing the benefits and flexibility of RUCA.
For 3-level designs, RUCA saves power consumption by 36.57\% within 1\% error and by 51.32\% within 2\% error on average.
\end{abstract}


\section{Introduction}
\label{sec:intro}

As circuit customization is developed to meet the requirements of various applications, power consumption becomes a main factor limiting the scale of computational capacity. Approximate computing is one of the emerging low-power techniques, which aims to improve power consumption as well as circuit delay by relaxing the requirement for 100\% accuracy. Approximate computing can be widely used in many application domains, such as machine learning, computer vision and signal processing, which have inherent resilience to small inaccuracies in the outputs~\cite{han2013approximate}. Such resilience can originate from various sources including, noise in input data, inherent approximate calculations, or human tolerance to variations in the outputs, while different applications may have different resilience. Thus, one challenge of approximate computing is to design approximate circuits which are able to dynamically switch among various accuracy levels (including full accuracy) at runtime, each of which is associated with different power consumption.  By properly configuring accuracy levels at runtime, power consumption could be substantially saved.


The last few years have seen various techniques for approximate logic synthesis~\cite{salsa, aslan14, blasys, Nepal14, Vasicek2016}. Most of them only generate ``fixed'' approximate circuits without the flexibility of runtime configuration. Meanwhile, some other works start to explore methodologies of runtime configurable circuits~\cite{zervakis2020design,venkataramani2013substitute,jain2016approximation,alan2020runtime,moons2015dvas}.
In this paper, we propose a novel {\it RUntime Configurable Approximation} (RUCA) methodology based on {\it factorizing} and {\it separating} truth tables. RUCA generates approximate circuit with multiple accuracy levels, including full accuracy when needed. The contributions of this paper are as follow.

\begin{itemize}
    \item Utilizing Boolean Matrix Factorization (BMF) algorithm, our novel RUCA approach {\it approximates} an arbitrary input circuit and {\it separates} it into multiple {\it configuration blocks} by decomposing factorized truth tables. 
    By enabling different blocks at runtime, we can dynamically choose the expected accuracy-power configuration, where enabling more blocks improves accuracy at the expense of power consumption.
    \item A corrector circuit is introduced to restore the functionality of the original correct circuit. With the corrector circuit, RUCA  is  able to operate under 100\% accuracy  when needed.
    \item To improve the scalability of our approach, a large input circuit is first partitioned into subcircuits with manageable size, and a design space exploration scheme is used to locate the proper subcircuits to approximate in the runtime configurable manner. Blocks of each subcircuit are then assigned to {\it configuration blocks}  for the \textit{top-level} circuit.
    \item We evaluate RUCA framework on a number of commonly used arithmetic circuits from Benchmarks for Approximate Circuit Synthesis (BACS)~\cite{BACS} and EPFL benchmark~\cite{amaru2015epfl}. We also compare our methodology against another quality-configurable framework, Approximate through Logic Isolation~\cite{jain2016approximation}, showcasing that our approach efficiently improves power utilization with the flexibility of accuracy-power configurations. For 3-level designs, on average RUCA saves power consumption by 36.57\% within 1\% error and by 51.32\% within 2\% error.
\end{itemize}

The organization of this paper is as follow. In Section~\ref{sec:prev_work}, we overview relevant previous work on approximate logic synthesis. 
In Section~\ref{sec:bmf}, we discuss the problem of Boolean Matrix Factorization and its application in approximate computing.
In Section~\ref{sec:methodology}, we introduce our novel RUCA methodology. We provide our experimental results in Section~\ref{sec:results}. Finally, we summarize our conclusion and directions for future works in Section~\ref{sec:conclusion}.

\section{Previous Work}
\label{sec:prev_work}

Recent study on approximate logic synthesis can be divided into two categories: Boolean or gate-level approaches and high-level synthesis approaches.

For Boolean or gate-level approaches, a number of methodologies have been proposed. In SALSA~\cite{salsa}, a miter is created to compute the error between the original circuit and the approximate circuit using the existing techniques in logic synthesis. The don't cares of the outputs of the approximate circuit with respect to outputs of the difference circuit can be used to simplify the approximate circuit using regular logic synthesis techniques. This approach was extended in ASLAN~\cite{aslan14} to model error arising over multiple cycles. 
In BLASYS~\cite{blasys}, an input circuit can be approximated by simulating into truth table, which is factorized into two smaller matrices and then synthesized into the approximate circuits.

For higher-level synthesis, ABACUS~\cite{Nepal14} generates variants of an input high-level Verilog description file by applying a set of possible transformations on the circuit to generate a set of mutant approximate circuit variants. A multi-objective design space exploration technique is then used to identify the best set of approximate variants. In EvoApprox~\cite{Vasicek2016}, a genetic algorithm is used to approximate arithmetic blocks, where the exact circuit is encoded in a string-based representation as a ``chromosome'' and mutated to create approximations as long as the error is kept below target.

Compared to general approximate computing methodologies, runtime configurable design is less explored. One category of runtime configuration is voltage over-scaling (VOS), where the power and accuracy of operation can be dynamically adjusted by tuning the voltage. However, the application of VOS is limited since it may cause uncontrollable errors that potentially affect the most significant bits. Also, VOS increases delays on all timing paths, which may affect the performance of the whole system and even lead to the failure of operation~\cite{reda2019approximate}.

In order to design stable and predictable runtime configurable circuits, few methodologies based on approximate logic synthesis have been proposed~\cite{kahng2012accuracy,albicocco2013truncated,mrazek2018design}. An accuracy-configurable approximate adder was proposed~\cite{kahng2012accuracy}, where input operands are split into multiple segments, and each segment is summed up independently in order to compose different accuracy levels. Power consumption can be saved by disabling summation of some segments. 
For multipliers, a programmable truncated multiplier was proposed~\cite{albicocco2013truncated}, which aims to disable less-significant columns of multiplier using power-gating. Another framework for dynamically configurable multipliers was also proposed based on Cartesian Genetic Programming~\cite{mrazek2018design}.

SASIMI~\cite{venkataramani2013substitute} proposed the first methodology to generate accuracy configurable design from an arbitrary input circuit by identifying similar signals and substituting one for the other to simplify the logic.  However, when full accuracy is required, the approximate circuit may need an additional clock cycle to retain the original signal and re-compute the accuracy outputs. Although energy is saved, SASIMI turns a combinational circuit into a variable latency circuit, which may not be applicable to large systems. 

\begin{figure}[t!]
	\begin{center}
    \includegraphics[width=0.45\textwidth]{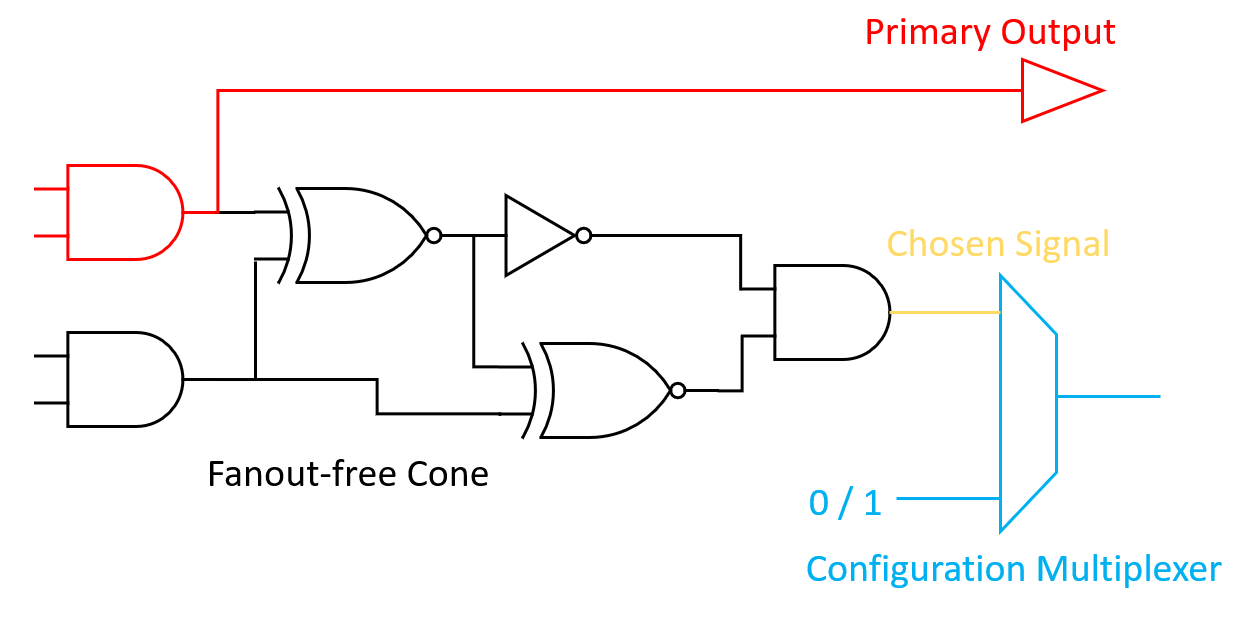}

		\caption{Implementation of Approximation through Logic Isolation. The yellow line represents selected signal. The blue node represents the multiplexer for signal configuration. Black nodes consist a fanout-free cone. Notice that the red node is not part of the fanout-free cone, since there exists another path between red node and primary outputs. }
		\label{fig:Iso}
	\end{center}
\end{figure}

To mitigate the possibly doubled delay, an approximation approach through logic isolation is proposed~\cite{jain2016approximation}, which aims to isolate portions of logic that significantly contribute to power consumption, but have less effect on overall accuracy. By identifying the proper signals and disabling the fanout-free cones, the power consumption of overall circuit is reduced by trading off a limited amount of accuracy.
In Section~\ref{sec:results}, we implement Approximation through Logic Isolation as comparison against our methodology. We analyze the relationship between each signal of the original circuit and final outputs. Specifically, each signal is substituted by a fixed value of 0 or 1, and the errors on final output are
measured as the effect of that signal on Quality-of-Results (QoR). Each signal is also associated with a set of {\it fanout-free cones}, where the paths between each node of fanout-free cone to any primary output must include that signal. Thus, disabling a fanout-free cone is equivalent to fix the value on the corresponding signal. We choose the {\it best} fanout-free cone, which has largest power consumption. By trading off the effect on QoR and power saving of disabling the {\it best} fanout-free cone, we greedily select signals that save more power while minimizing errors on final outputs. A multiplexer is inserted at the chosen signal as configuration between original value and fixed value. When the original value is expected for the signal, the fanout-free cone is activated and the multiplexer is configured to choose the original source of signal. On the other hand, when a fixed value is expected for the purpose of approximation mode, the multiplexer is configured to choose the fixed value, and fanout-free cone is disabled to save power. By configuring these multiplexers, Approximation through Logic Isolation is able to achieve difference accuracy levels. Figure~\ref{fig:Iso} demonstrates an example of signal configuration and fanout-free cone.

Another method based on logic gating is also proposed~\cite{alan2020runtime}, where multiple approximate designs of input circuit are first instantiated. Then area-saving gating mechanisms are used to exploit synthesis relaxation, which leads to total energy saving. While this methodology reduces power consumption significantly, a large amount of area overhead is introduced. In RETSINA~\cite{zervakis2020design}, simulated annealing is used to produce
accuracy configurable circuits by
combining gate-level pruning and wire-by-switch replacement.


\section{Background}
\label{sec:bmf}

In this section, we describe the problem of Boolean Matrix Factorization (BMF), as it forms the mathematical basis of our methodology. We also discuss the general methodology of approximate logic synthesis using BMF.

\begin{figure}[t]
	\begin{center}
		\includegraphics[width=\linewidth]{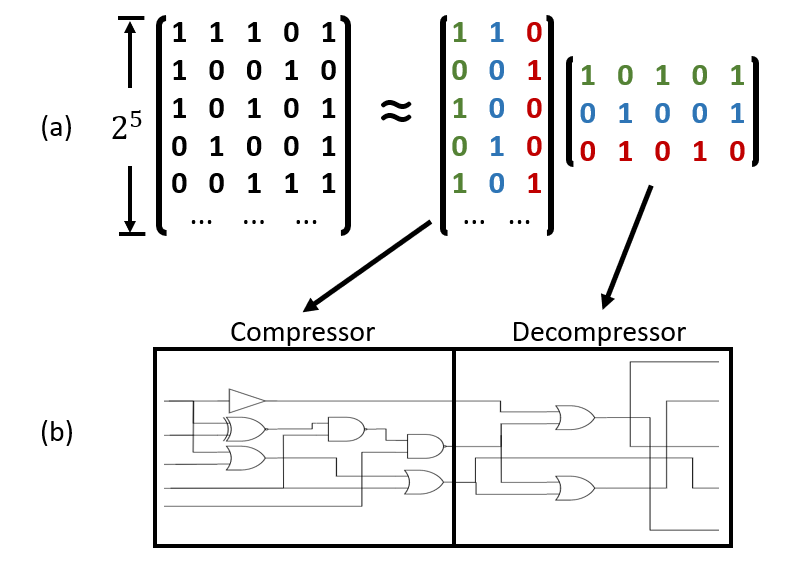}
		\vspace{-0.1in}
		\caption{(a) An example of Boolean matrix factorization, where green values are derived at first iteration, then blue ones at second iteration, finally red ones at third iteration. (b) An example of approximate logic synthesis using BMF.}
		\label{fig:heuristic}
	\end{center}
\end{figure}

A Boolean matrix is a special matrix where  all elements are limited to Boolean values, \textit{i.e.}, `0's or `1's.  Boolean Matrix Factorization aims to factorize an input Boolean matrix $\mathbf{M}$ of size $p \times q$ into two Boolean matrices: a $p \times f$ matrix, $\mathbf{A}$, and a  $f \times q$ matrix, $\mathbf{B}$, such that $\mathbf{M} \approx \mathbf{A}\mathbf{B}$, where $f$ is called \textit{factorization degree}. In many applications, factorization degree $f$ is required to be smaller than $q$.
The multiplications are carried out using the logical AND operation, while the additions can be performed by logical OR operation~\cite{Miettinen11}. Note that one can interpret the columns of $\mathbf{A}$ as {\it basis} vectors, which are linearly combined using $\mathbf{B}$. BMF has been proved to be NP-hard~\cite{miettinen2008discrete}, which can also be formulated as an optimization problem to minimize errors resulted from factorization,

\begin{equation}
\text{argmin}_{\mathbf{A},\mathbf{B}}|\mathbf{M} - \mathbf{A} \mathbf{B}|
\label{eq:BMF}
\end{equation}
where the elements of $\mathbf{M}$, $\mathbf{A}$ and $\mathbf{B}$ are Boolean matrices. Due to its NP-hardness, many algorithms solve BMF using heuristic approaches. In our approach, we considered an algorithm based on association rule mining (ASSO)~\cite{miettinen2008discrete}. To begin with, an association matrix $\mathbf{O}$ is computed, where each row is considered as a candidate of {\it basis} vectors in $\mathbf{B}$.
For each candidate, ASSO computes a paired column vector by exhaustive search. With \textit{factorization degree} $f$, ASSO greedily picks $f$ pairs of row and column \textit{one-by-one} in order to cover as many `1's as possible in input matrix $\mathbf{M}$.
The heuristic property indicates that at each iteration from 1 to $f$, the factorization result is always {\it locally optimal}.
For an input matrix of size $p\times q$, the time complexity of ASSO algorithm is $O(pq^2)$. Figure~\ref{fig:heuristic} demonstrates an example of factoring a $2^5 \times 5$ input matrix with factorization degree $3$. The process mentioned before is carried out for three iterations, solving $3$ pairs of columns and rows {\it one-by-one}. At each iteration, results from previous iterations are amended with an additional pair of column and row, where $|\mathbf{M} - \mathbf{A} \mathbf{B}|$ keeps reducing after each amendment.


There exists an inherent connection between logic circuits and BMF, where truth tables of combinational logic are effectively Boolean matrices. Thus, as proposed in BLASYS~\cite{blasys}, BMF can be used to approximate an arbitrary circuit with $n$ inputs and $m$ outputs. The exact input circuit is simulated to obtain the truth table $\textbf{M}$ of size $2^n \times m$, which is then given as input to a BMF algorithm together with a factorization degree $1\leq f < m$. $\textbf{M}$ is then factorized into a $2^n \times f$ matrix $\textbf{A}$, and a $f \times m$ matrix $\textbf{B}$. Figure~\ref{fig:heuristic} illustrates an example of approximate logic synthesis using BMF, where $n=5$, $m=5$ and $f=3$. Using existing logic synthesis techniques, first Boolean matrix $\textbf{A}$ is used to synthesize the first part of approximate circuit with $n$ inputs and $f$ outputs, which is referred to as {\it compressor} circuit. The second part receives $f < m$ inputs from {\it compressor} circuit and maps them back to $m$ outputs.
This subcircuit is referred to as {\it decompressor} circuit, which can be generated using a network of OR gates according to each column in $\textbf{B}$. 
Compared to other gate-level approximate computing methodologies, {\it e.g.} SASIMI~\cite{venkataramani2013substitute}, BMF-based methodology has stronger control over the error introduced by approximation. By gradually changing the factorization degree, we are able to obtain numerous approximate designs with different accuracy levels. 


\section{Proposed Methodology}
\label{sec:methodology}

In this section, we describe our proposed methodology of designing runtime configurable approximate circuit by factorizing and separating truth table, together with the method of self-correcting by corrector circuit. Due to the complexity of BMF algorithm, a divide-and-conquer approach is considered to improve the scalability of our methodology, where we proposed to use circuit partitioning and design space exploration (DSE) scheme to apply our framework on large circuits.

\subsection{RUntime Configurable Accuracy (RUCA) with Corrector Circuit}
\label{sec:runtime}

According to the rule of matrix multiplication, after factorizing a matrix $\mathbf{M}$ into $\mathbf{A}$ and $\mathbf{B}$, we may separate them into individual columns and rows, as Equation~\ref{eq:factor1},

\begin{equation}
\footnotesize 
    \mathbf{M} \approx \mathbf{A}\mathbf{B}=(\mathbf{a_1} \cdots \mathbf{a_f}) \left(\begin{array}{c}
\mathbf{b_1}\\
\vdots\\
\mathbf{b_f}
\end{array}\right)= \mathbf{a_1}\mathbf{b_1} + \mathbf{a_2}\mathbf{b_2} + \cdots+ \mathbf{a_f}\mathbf{b_f}
\label{eq:factor1}
\end{equation}
\noindent where $\mathbf{a_i}$ is the $i^{th}$ column in matrix $\mathbf{A}$ and $\mathbf{b_j}$ is  $j^{th}$ row of matrix $\mathbf{B}$. As discussed in Section~\ref{sec:bmf}, 
due to the heuristic property of BMF algorithm, as we add terms from $\mathbf{a_1}\mathbf{b_1}$ to $\mathbf{a_f}\mathbf{b_f}$, the difference between factorized and original matrix $|\mathbf{M}-\mathbf{A}\mathbf{B}|$ keeps decreasing in a {\it greedy} manner.
To implement runtime configuration, our goal is to factorize the input matrix $\mathbf{M}$ with multiple error thresholds.  For example, suppose that we want to factorize $\mathbf{M}$ such that there exists two configurable error thresholds (e.g., 2\% and 1\%). Starting from factorization degree $f=1$ with only the first term $\mathbf{a_1}\mathbf{b_1}$, we gradually increment $f$ and sum $\mathbf{a_i}\mathbf{b_i}$ terms, until the QoR difference between  $\mathbf{M}$ and $\mathbf{A}\mathbf{B}$ becomes no larger than 2\%. Assume the current factorization degree is $f=k_1$. In this case, we can stack vectors from $\mathbf{a_1}$ to $\mathbf{a_{k_1}}$ as $\mathbf{A_1}$, and stack vectors from $\mathbf{b_1}$ to $\mathbf{b_{k_1}}$ as $\mathbf{B_1}$, such that the QoR difference between  $\mathbf{M}$ and $\mathbf{A_1}\mathbf{B_1}$ is no larger than $2\%$.  We then keep incrementing $f$ until 1\% error threshold  is met. Assuming now $f=k_1+k_2$,  vectors from $\mathbf{a_{k_1+1}}$ to $\mathbf{a_{k_2}}$ are stacked as $\mathbf{A_2}$, and vectors from $\mathbf{b_{k_1+1}}$ to $\mathbf{b_{k_2}}$ are stacked as $\mathbf{B_2}$, such that the QoR difference between $\mathbf{M}$  and $\mathbf{A_1}\mathbf{B_1}+\mathbf{A_2}\mathbf{B_2}$ is no greater than $1\%$.
In other words, we propose to separate factorized matrices $\mathbf{A}$ and $\mathbf{B}$,  such that

\begin{figure}[t!]
	\begin{center}
	
	\subfloat[][]{
    \includegraphics[width=0.5\textwidth]{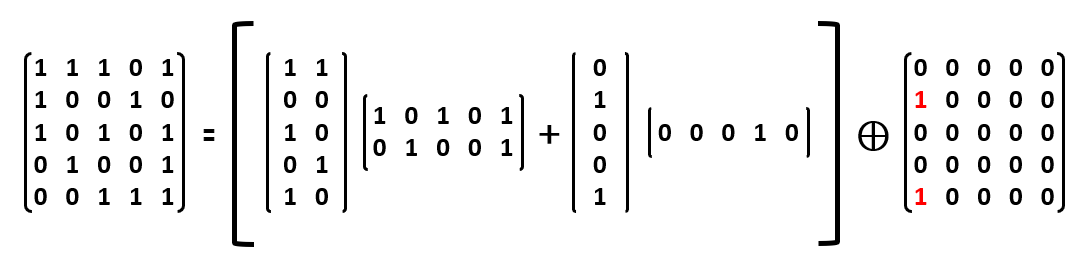}
    \label{fig:BMF_level}}
    \\
	\subfloat[][]{
    \includegraphics[width=0.5\textwidth]{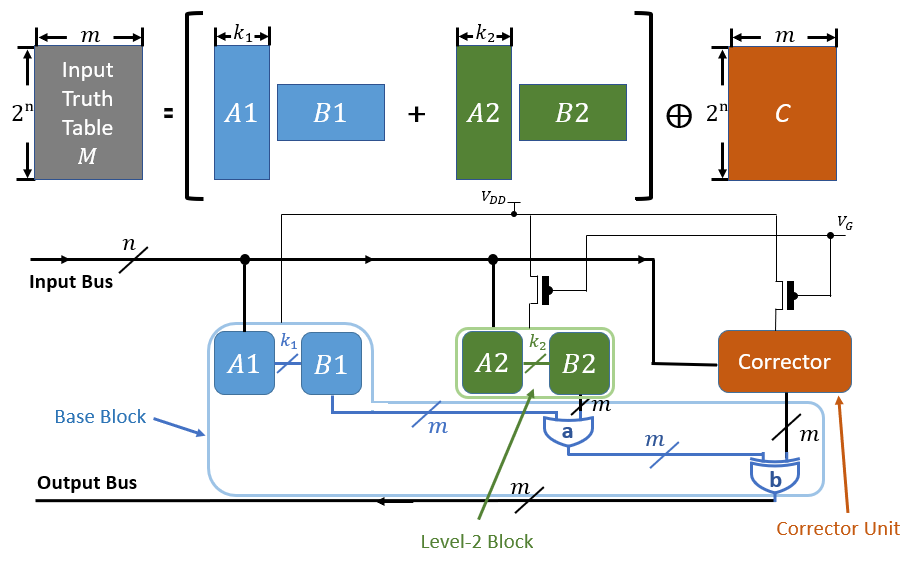}
    \label{fig:circuit_level}}
		\caption{Example of a 3-level approximate circuits using RUCA. (a) BMF with multiple accuracy levels. (b) Runtime configurable circuit design, where power gating is used to activate different blocks. Gate $a$ represents bitwise OR between outputs of the factor circuits. Gate $b$ represents bitwise XOR between approximate and corrector outputs.}
		\label{fig:general_flow}
	\end{center}
\end{figure}
{\footnotesize
\begin{align}
    \mathbf{M} & \approx \mathbf{A}\mathbf{B}= \mathbf{A_1}\mathbf{B_1}+\mathbf{A_2}\mathbf{B_2} \nonumber \\
    & =(\mathbf{a_1} \cdots \mathbf{a_{k_1}}) \left(\begin{array}{c}
\mathbf{b_1}\\
\vdots\\
\mathbf{b_{k_1}}
\end{array}\right)+ (\mathbf{a_{k_1+1}} \cdots \mathbf{a_f}) \left(\begin{array}{c}
\mathbf{b_{k_1+1}}\\
\vdots\\
\mathbf{b_f}
\end{array}\right)
\label{eq:factor}
\end{align}
}

\begin{figure*}[t!]
 	\begin{center}
 		\includegraphics[width=0.87\linewidth]{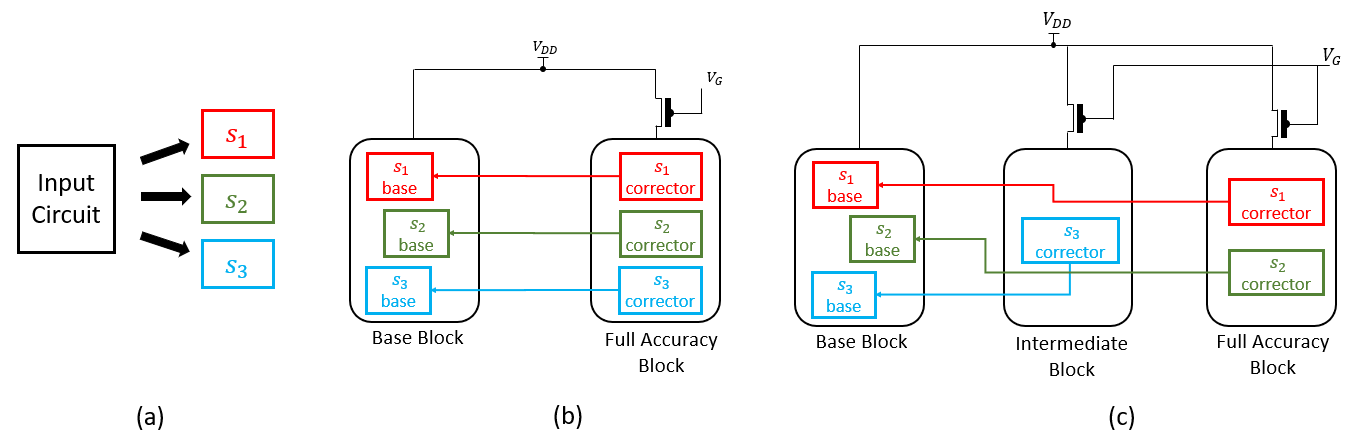}
 		\caption{An example of runtime configurable designs for a large input circuit. (a) Input circuit is partitioned into three subcircuits. (b) {\it Subcircuits} are approximated into 2-level runtime configurable designs, each with a base block and a corrector circuit. Then 3 base blocks of {\it subcircuits} are synthesized together as the base block of the {\it top-level circuit}. 3 corrector circuits are grouped together as the full-accuracy block of the {\it top-level circuit}. (c) Additional accuracy levels can be introduced by re-arranging corrector circuits of {\it subcircuits} into intermediate block(s). }
 		\label{fig:method}
 	\end{center}
 \end{figure*}

If more accuracy levels are needed, this procedure is repeated until we obtain matrices $\mathbf{A_i}$ and $\mathbf{B_i}$ for each accuracy level or factorization degree reaches $f=m-1$, where $m$ is the number of primary outputs in the given circuit. We propose to synthesize each $\mathbf{A_i}\mathbf{B_i}$ term into its own block as shown in Figure~\ref{fig:general_flow}. To implement binary addition, bitwise OR gates are used to connect each block of $\mathbf{A_i}\mathbf{B_i}$ term. Therefore, starting from block of $\mathbf{A_1}\mathbf{B_1}$, as we activate more $\mathbf{A_i}\mathbf{B_i}$ blocks, the difference between original truth table $\mathbf{M}$ and summation of truth tables of $\mathbf{A_i}\mathbf{B_i}$ keeps decreasing, where different error thresholds can be achieved.

In order to support critical applications which require full accuracy, we propose to use a corrector circuit to restore the original functionality when needed. Here, field modulo-2 algebra (logic XOR) is used to correct flipped bits, where `1's can be used to flip bits such that $1\oplus 1=0$ and $1\oplus 0=1$. After input truth table $\mathbf{M}$ is factorized into $\mathbf{A}$ and $\mathbf{B}$, bitwise XOR is computed between $\mathbf{M}$ and approximate truth table $\mathbf{A}\mathbf{B}$ to obtain the corrector matrix $\mathbf{C}$. This matrix can be used to restore input truth table $\mathbf{M}$ by $\mathbf{M}=\mathbf{A}\mathbf{B}\oplus \mathbf{C}$. 

Figure~\ref{fig:BMF_level} demonstrates a factorization algebra with three accuracy levels. The rightmost matrix is the corrector matrix $\textbf{C}$, which is computed to restore the input matrix by XOR operation. Figure~\ref{fig:circuit_level} demonstrates structure of a 3-level runtime configurable circuit. Firstly, an input circuit with $n$ inputs and $m$ outputs is simulated to obtain the $2^n\times m$ truth table $\textbf{M}$. Then, $\textbf{M}$ is  factorized into two matrices $\mathbf{A}$ and $\mathbf{B}$, which are then separated into $\mathbf{A_1}\mathbf{B_1}$ and $\mathbf{A_2}\mathbf{B_2}$, according to previous discussion. All matrices are synthesized into corresponding parts of the circuit. Corrector matrix $\mathbf{C}$, which is used to synthesize the corrector circuit, is computed for restoring input truth table $\mathbf{M}$. As Boolean algebra indicates, $\mathbf{A_1}\mathbf{B_1}$ and $\mathbf{A_2}\mathbf{B_2}$ are connected using bitwise OR gates $a$, which is then connected to the corrector circuit using bitwise XOR gates $b$. Thus, if all parts are activated, it will produce equivalent functionality as original circuit, where circuit runs in full-accuracy mode:
\begin{equation}
    \mathbf{M} = (\mathbf{A_1}\mathbf{B_1} + \mathbf{A_2}\mathbf{B_2}) \oplus \mathbf{C} 
\end{equation}
In order to enable runtime configuration, we combine these parts into different {\it configurable blocks}, and use power gating to control their activation. In this example, $\mathbf{A_1}$, $\mathbf{B_1}$ and all connecting gates compose the \textit{base} block, which is always activated by default. When the \textit{base} block is the only enabled one, the circuit operates in approximate mode with lowest accuracy, where the output 
matrix $\mathbf{M'}$ is
\begin{equation}
    \mathbf{M'} = \mathbf{A_1}\mathbf{B_1}
\end{equation}
$\mathbf{A_2}$ and $\mathbf{B_2}$ compose the \textit{level-2} block. For higher-accuracy approximate mode,  \textit{level-2} block is additionally activated, where the output matrix $\mathbf{M''}$ is
\begin{equation}
    \mathbf{M''} = \mathbf{A_1}\mathbf{B_1} + \mathbf{A_2}\mathbf{B_2}
\end{equation}

\noindent Following this framework, we are able to design runtime configurable circuits with arbitrary number of accuracy levels.

Design overhead is considered as an important criterion in runtime configurable designs, which is defined as additional chip area and power consumption running in full-accuracy level compared to the input circuit. According to Figure~\ref{fig:general_flow}, the configuration overhead of our design comes from the connecting OR and XOR gates. Also, since blocks for different accuracy levels are synthesized separately, we may lose the opportunity of logic optimization across different blocks. Thus, design overhead also comes from the logic redundancy between each accuracy level. In section~\ref{sec:results}, we analyze the trade-off between design overhead and choices of error thresholds.

\subsection{Partitioning and Design Space Exploration}
\label{sec:DSE}
 
The number of rows in a truth table grows exponentially with the number of primary inputs in the circuit, which makes the BMF factorization algorithm computationally expensive for large circuits. 
To scale our approach, we propose to adopt \textit{divide-and-conquer} using circuit partitioning and design space exploration technique.
To begin with, a given circuit is partitioned into a number of subcircuits with manageable size, each of which is approximated using our runtime configurable approach as illustrated in Figure~\ref{fig:method}. A design space exploration technique is used to navigate the search space to find proper subcircuits and factorization degrees as described in Algorithm~\ref{alg:dse}.
The design overhead of our framework mainly comes from (1) the connecting OR and XOR gates, and (2) the logic redundancy between each accuracy level. In other words, the design overhead increases as the number of accuracy levels increases. Figure \ref{fig:method} illustrates an example, where each  subcircuit is approximated with two levels, which only consists of a base block and a corrector circuit.
As illustrated in Figure~\ref{fig:method}b, base blocks of the approximate subcircuits are grouped in an individual power domain and synthesized together as the base block of the  top-level design. During this process, logic optimization is performed for all base blocks of subcircuits, which helps remove logic redundancy. Corrector circuits of approximate subcircuits are also grouped together  as full-accuracy block as illustrated in Figure \ref{fig:method}b, which enables the top-level design to restore full accuracy. Moreover, if additional accuracy levels are expected, more intermediate blocks can be created by re-arranging the corrector circuits of subcircuits, as illustrated in Figure~\ref{fig:method}c.

\setlength{\textfloatsep}{4pt}
\begin{algorithm}[t!]
	\small
	\SetKwBlock{Begin}{begin}{end}
	\SetKwInOut{Input}{Input}
	\SetKwInOut{Output}{Output}
	\Input{Input Circuit $ICir$, List of Error Thresholds $\epsilon$ in ascending order}\label{alg:input}
	\Output{Runtime configurable circuit $Cir$ with self correcting capability}
	$S_{Cir}$ = Partition $ICir$ into subcircuits with maximum $n$ inputs and $m$ outputs\\
	// Factorize truth table for each subcircuit \\
	\For{each subcircuit $s_i$ with $m_i$ outputs}
    {
     $\textbf{M}_i$ = Simulate subcircuit $s_i$ \\
    $[\textbf{A}_i, \textbf{B}_i]$ = BMF($\textbf{M}_i$, $\, f = m_i-1$) 
    }
    // Begin Design Space Exploration \\
    $Cir$ = $ICir$ \\
    $A_{Cir}$ = $\varnothing$  // Set of approximated subcircuits \\
    Let $f_i$ = $m_i$ for each $s_i$ in $S_{Cir}$ \\
    \While{$\epsilon$ is not empty}
    {
        \For{each subcircuit $s_i$ in $S_{Cir}$}
        {
            $Cir_i$ = RUCA($s_i,f_i-1$) \\
            $QoR_i$ = Error of design $Cir_i$ \\
            $P_{acc}(Cir_i)$ = Power under full-accuracy mode \\
            $P_{app}(Cir_i)$ = Power under approximation mode \\
            $loss_i$ = $QoR_i\cdot [P_{acc}(Cir_i)+P_{app}(Cir_i)]$ \\
        }
        $k$ = $\arg\min_i\, loss_i$ \\
        $A_{Cir}$ = $A_{Cir} \cup s_k$ \\
        \If{$QoR_k \geq  \, \epsilon[0]$}
        {
        \For{each subcircuit $s_i$ in $A_{Cir}$}
        {
        $Cir\leftarrow$  RUCA($s_{i},f_{i})$ \\
        $S_{Cir}$.pop($s_i$)\\
        }
        $\epsilon$.pop(0) \\
        }
        $f_k$ = $f_k$ - 1 \\
    
    }	
	\Return $Cir$ 
	\caption{Runtime Configurable Approximate Circuit with Design Space Exploration}
		\label{alg:dse}
\end{algorithm}
 
Algorithm~\ref{alg:dse} describes the overall procedure. To begin with, the input circuit is partitioned into subcircuits (line 1), where hypergraph partitioning algorithm~\cite{shhmss2016alenex} is executed recursively, such that subcircuit $s_i$ has $m_i$ outputs (line 3). Since we want to efficiently factorize the truth table of each subcircuit, the number of inputs and outputs of each subcircuit should be restricted, \textit{e.g.} $n_1\leq10$ and $m_i\leq10$. For each subcircuit $s_i$, the truth table $\mathbf{M_i}$ is obtained and then factorized into $\mathbf{A_i}$ and $\mathbf{B_i}$ with factorization degree $m_i-1$ (line 5).

In our design space exploration scheme (Lines 8-29), we gradually increase level of approximation by substituting subcircuit with a new 2-level runtime configurable design. For each subcircuit $s_i$, factorization degree $f_i$ is searched from $m_i-1$ to $1$, where proper degrees will be used to split truth tables as discussed in section ~\ref{sec:runtime}. At each iteration, we go through each subcircuit that has not been approximated (line 12), whose current factorization degree $f_i$ is decreased by 1 and then used to generate a new runtime configurable design (line 13), where RUCA$(s_i,f_i-1)$ denotes a runtime configurable design based on subcircuit $s_i$ with \textit{factorization degree} $f_i-1$.
In each iteration, we generate a new runtime configurable design by approximating each subcircuit {\it individually} as candidate designs. For each candidiate design, we evaluate the Quality of Results $QoR_i$ by error, power consumption when running in full-accuracy mode $P_{acc}(Cir_i)$, and power consumption when running in approximation mode $P_{app}(Cir_i)$. And a loss value (line 17) is computed to minimize power consumption in both full-accuracy mode and approximate mode, while the errors in approximate mode are also expected to be low. Among all candidate designs, we find the one with least loss value (line 19). At this point, if current error reaches the smallest threshold in the list $\epsilon$ (line 21), all modifications {\it until previous iteration} are updated to current circuit $Cir$ (line 23) as starting point of following iterations. As mentioned previously, in order to limit the design overhead, each subcircuit is approximated with only \textit{two} levels. Thus, we also remove already-approximated subcircuits from candidate list (line 24). Before next iteration, we also need to decrease the factorization degree of corresponding subcircuit by 1 (line 28).  The previous process is repeated until we create configurable blocks for all error thresholds in $\epsilon$. The accuracy level of top-level circuit can be controlled by power-gating different configurable blocks such as those in Figure~\ref{fig:method}.

By limiting the number of inputs $n$ and number of outputs $m$ in each subcircuit, out approach is able to efficiently work on large input circuits. Time complexity of each iteration is $O(m^2n^2|S|)$, where $|S|$ denotes the number of subcircuits. However, since we limit  $n\leq 10$ and $m\leq 10$, it is effectively linear to $i$. As local optimum is chosen at each iteration, theoretically speaking, our algorithm does not guarantee global optimal design. But in practice, since a large input circuit is partitioned into a number of small subcircuits, our methodology explores a large amount of possible approximations. As shown in section~\ref{sec:results}, our methodology is able to generate promising runtime configurable designs.

\subsection{Reducing  Design Overhead}
\label{sec:overhead}

In our design, the configurable overhead mainly comes from (1) the connecting OR and XOR gates, and (2) the logic redundancy between each accuracy level. These connecting gates are inevitable in our framework. To mitigate such issue, we have restricted the approximated subcircuits to {\it two levels}. In this subsection, we mainly focus on the issue of logic redundancy, especially from the perspective of corrector circuit.

\begin{figure}[b!]
 	\begin{center}
 		\includegraphics[width=1\linewidth]{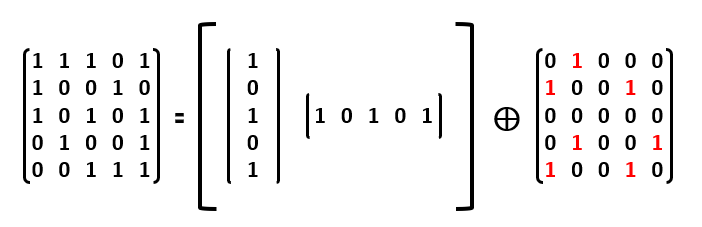}
 		\caption{Corrector matrix becomes less sparse when factorization degree is low.}
 		\label{fig:less-sparse}
 	\end{center}
 \end{figure}

As discussed in subsection~\ref{sec:runtime}, the corrector circuit is synthesized from corrector truth table, which flips wrong bits in approximated truth table. Normally, the difference between approximated and original truth table is not too large, when the matrix corresponding to corrector circuit  is sparse as shown in Figure~\ref{fig:BMF_level}. In this case, the overhead caused by corrector circuit is small. However, if input circuit is partitioned and design space exploration is performed, some subcircuits may be approximated to a low factorization degree, where the difference between approximated and original truth table is quite large, as Figure~\ref{fig:less-sparse} shows. In this situation, we can end up having a significantly large corrector circuit, sometimes even larger than the original subcircuit. In this case, rather than using a corrector circuit to achieve full accuracy, we use the original subcircuit instead. In our design space exploration algorithm, once the corrector circuit is synthesized, we compare the power consumption between corrector circuit and original one. If a corrector circuit consumes less power, we follow the algorithm described in Section~\ref{sec:DSE}. However, if the corrector circuit consumes more power than the original subcircuit, we directly include original subcircuit for full-accuracy mode. In this case,  instead of XOR gates, a multiplexer is used to connect original subcircuit with the approximate versions.

\section{Experimental Results}
\label{sec:results}
\begin{table}[b!]
  \scriptsize
  \centering
    \caption{Characteristics of evaluated benchmarks.}
  \begin{tabular}{||l|l|l||l|l|l||}
\hline
Bench- & \multirow{2}{*}{Name} & \multirow{2}{*}{Function}    & \multirow{2}{*}{I/O} & Area & Power  \\
 mark &  &    &  & ($um^2$) & ($uW$)  \\
 \hline \hline
\multirow{7}{*}{BACS} & adder8 & 8-bit adder & 16/9 & 47.58 & 24.70 \\ 
& abs\_diff & absolute difference   & 16/9 & 67.41 & 22.68  \\ 
& adder32 & 32-bit adder    & 64/33  &  167.03 & 32.20 \\
& buttfly & butterfly structure   & 32/34 & 174.26  & 42.30  \\
& mac& multiply-add     & 12/8 & 94.48 & 33.76 \\ 
& mult8 & 8-bit multiplier   & 16/16 & 364.61 & 82.21  \\ 
& mult16 & 16-bit multiplier & 32/32 & 1084.52 & 245.06  \\ \hline
\multirow{3}{*}{EPFL} & bar & 128-bit barrel shifter & 135/128 & 3566.70 & 782.04 \\
 & max & 4-to-1 128-bit max & 512/130 & 4491.95 & 417.96 \\
 & sin & 24-bit sine & 24/25 & 7405.94 & 625.56  \\

 \hline
  \end{tabular}
  \label{table:information}
\end{table}

\begin{figure}[b!]
	\begin{center}
    \includegraphics[width=0.47\textwidth]{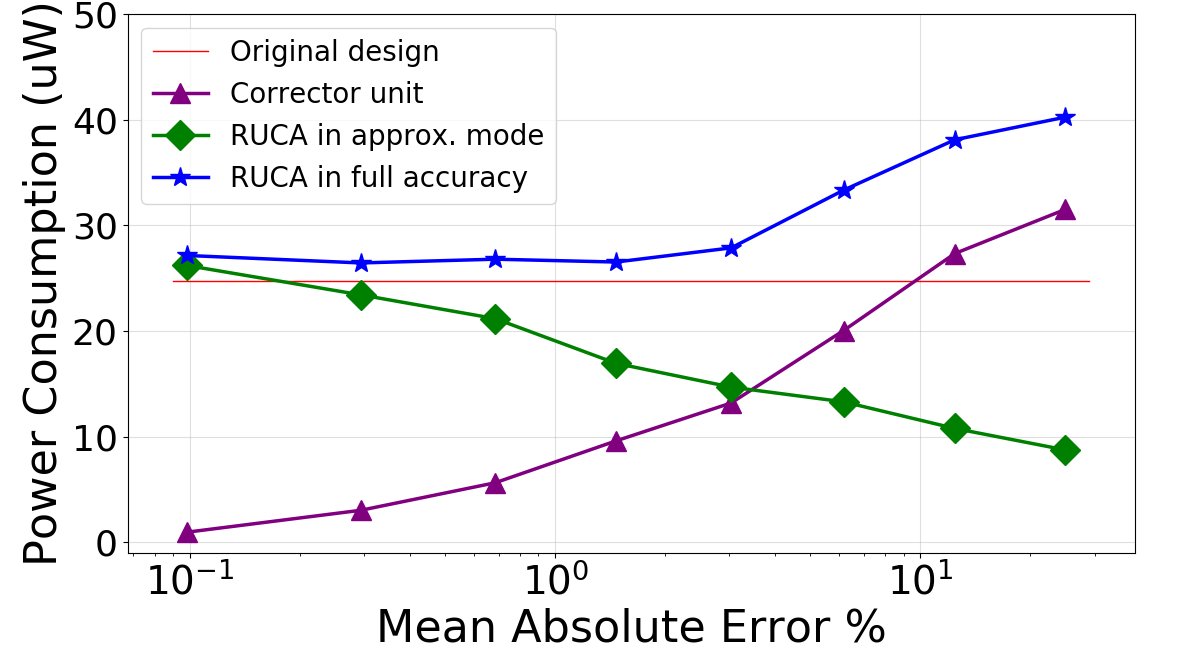}

		\caption{2-level approximate design of 8-bit adder: Power consumption with different error thresholds.}
		\label{fig:error}
	\end{center}
\end{figure}

\begin{figure*}[t!]
	\begin{center}

    \includegraphics[width=0.95\textwidth]{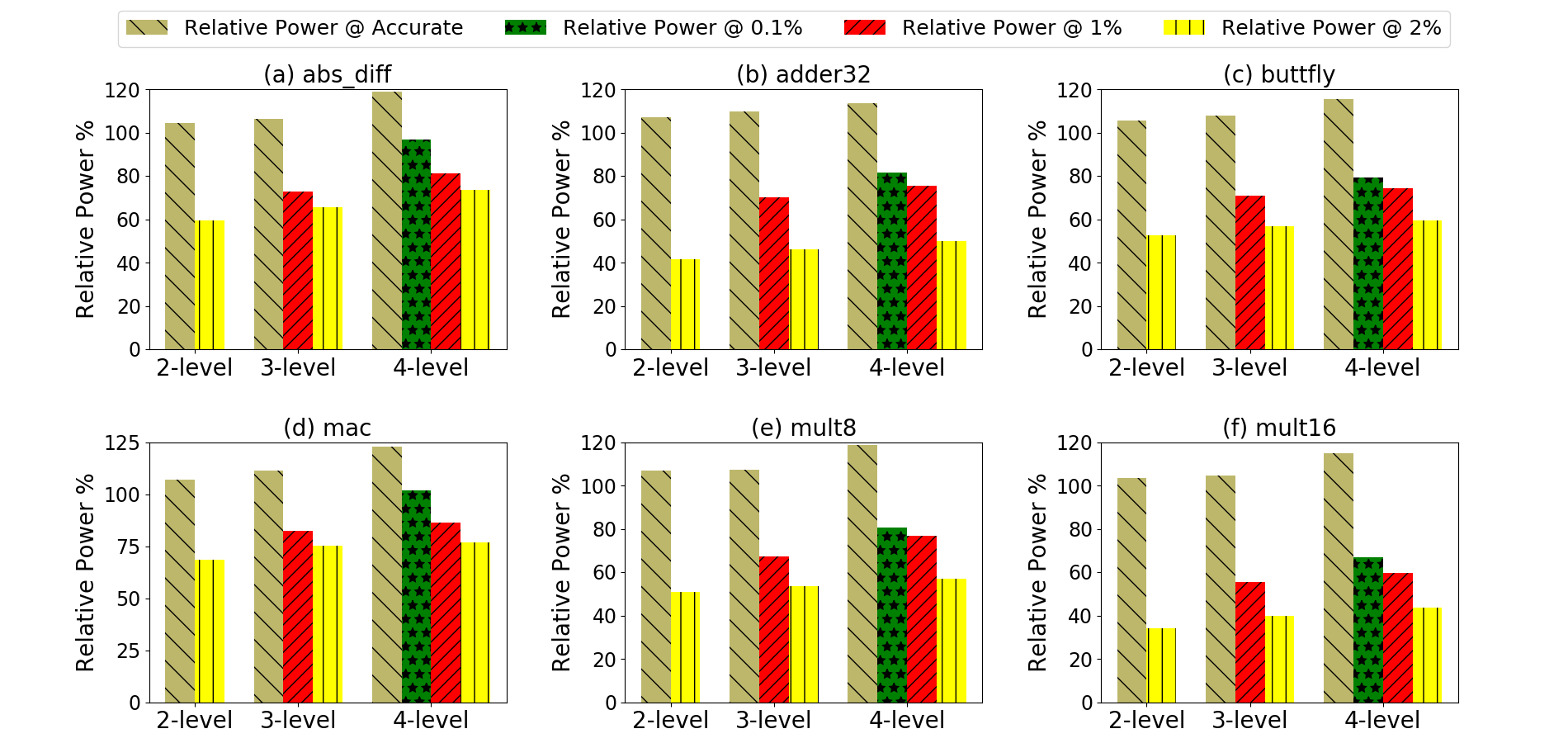}

		\caption{Relative power of RUCAs for each benchmark, with 2-4 levels.}
		
		\label{fig:result_power}
	\end{center}
\end{figure*}

In this section, we evaluate our proposed methodology on a number of arithmetic circuits deployed in approximate computing from Benchmarks for Approximate Circuit Synthesis (BACS)~\cite{BACS}. 
We also include three commonly used benchmarks from EPFL arithmetic benchmark suite~\cite{amaru2015epfl} to demonstrate the scalability.
Table~\ref{table:information} summarizes the characteristics of evaluated benchmarks. 
To begin with, we directly generate runtime configurable designs of 8-bit adder, where the trade-off between design overhead and choices of error thresholds is discussed.
The remaining benchmarks are first partitioned into subcircuits, and then design space exploration is performed as Algorithm~\ref{alg:dse}.

For hardware metrics, all designs are implemented in Verilog and synthesized with a $7nm$ predictive process design kit. Cadence Genus is used to synthesize each design and estimate chip area, circuit delay and power consumption under the maximum clock frequency of original circuit. For QoR metric, we report normalized mean absolute error (MAE) defined as
\begin{equation}
\mbox{MAE} = \frac{1}{N}\Sigma_{i=1}^{N}\frac{|R_{i}-R'_{i}|}{2^{m}},
\end{equation}
where $N$ denotes the size of the test vectors while $R_i$ and $R'_i$ denote the accurate and approximate numerical results.



In the first set of experiment, we analyze the trade-off between design accuracy, power consumption and design overhead.
Runtime configurable approximate circuits (RUCA) are generated for 8-bit adder with different error thresholds. Besides full accuracy, only {\it one} approximate level is considered for each design in this experiment. Since the original circuit has 9 primary outputs, after factorizing its truth table, first $f$ pairs of columns and rows are synthesized into \textit{base} block as approximate mode, where $f$ ranges from 1 to 8. For each RUCA design, an associated corrector circuit is created to restore errors in full-accuracy mode. In Figure~\ref{fig:error}, we report the power consumption of the corrector circuit, and the RUCA design in both approximate mode and full-accuracy mode. 
In approximate mode, power consumption reduces as error increases, where factorization degree $f$ is smaller. However, in full-accuracy mode, where the corrector circuit becomes more substantial and power-consuming. As MAE exceeds 5\%, where factorization degree $f < 4$, power consumption of full-accuracy mode increases substantially due to the corrector circuit. As Figure~\ref{fig:error} indicates, to limit the overhead in full-accuracy mode, error thresholds in approximate mode need to be limited, \textit{e.g.}, below 5\% MAE.

\begin{table*}[t!]
  \small
  \centering
    \caption{ Comparison of total area, relative power and circuit delay between RUCA and Approximation through Logic Isolation~\cite{jain2016approximation}  \qquad
    (using 3-level runtime configurable design) }
        \vspace{-0.1in}
  \begin{tabular}{||l|l||l|l|l||l|l||l|l||l|c||}
\hline
\multirow{3}{*}{Benchmark}     & \multirow{3}{*}{Name}     &  \multicolumn{3}{c||}{\multirow{2}{*}{Total Area ($nm^2$)}} & \multicolumn{2}{c||}{Relative Power} & \multicolumn{2}{c||}{Relative Power} & \multicolumn{2}{c||}{Relative Power} \\  
 & & \multicolumn{3}{c||}{} & \multicolumn{2}{c||}{under 2\% MAE} & \multicolumn{2}{c||}{under 1\% MAE} & \multicolumn{2}{c||}{under full accuracy} \\\cline{3-11}

 & &  RUCA & Isolation &  Saving \%  &  RUCA & Isolation & RUCA & Isolation & RUCA & Isolation   \\ \hline \hline
 
 \multirow{6}{*}{BACS} & abs\_diff &  \textcolor{red}{93.28}  &  115.96 & 19.56\% & \textcolor{blue}{65.79\%} & 55.82\% & \textcolor{red}{72.73\%} & 82.54\% & \textcolor{red}{105.68\%} & 109.85\% \\
 
 & adder32 &  \textcolor{blue}{238.42} & 224.60 & -6.15\% & \textcolor{red}{46.17\%} & 51.89\% & \textcolor{blue}{70.13\%}  & 60.93\% & \textcolor{red}{109.86\%} & 113.10\%  \\
 
 & buttfly &  \textcolor{blue}{257.44} & 237.16 & -8.55\% & \textcolor{blue}{56.84\%} & 51.82\% & \textcolor{red}{70.95\%} & 72.34\% & \textcolor{blue}{107.93\%} & 107.38\%  \\
 
 & mac &  \textcolor{red}{154.41} & 168.38 & 8.30\% & \textcolor{blue}{75.31\%} & 69.50\% & \textcolor{red}{82.39\%} & 83.38\% & \textcolor{red}{111.45\%} & 117.42\%  \\
 
 & mult8 &  \textcolor{red}{471.18} & 594.63 & 20.76\% & \textcolor{red}{53.70\%} & 71.29\% & \textcolor{red}{67.41\%} & 87.23\% & \textcolor{red}{107.41\%} & 112.49\%   \\
 
 & mult16 &  \textcolor{red}{1168.92} & 1395.42 & 16.23\% & \textcolor{red}{39.90\%} & 54.16\% & \textcolor{red}{55.47\%} & 72.45\% & \textcolor{red}{104.52\%} & 109.73\% \\ \hline

\multirow{3}{*}{EPFL} & bar & \textcolor{red}{4240.26} & 4738.88 & 10.52\% & \textcolor{blue}{22.40\%} & 17.62\% & \textcolor{blue}{41.69\%} & 39.73\%  & \textcolor{red}{109.27\%} & 114.92\% \\
& max  & \textcolor{red}{4917.92} & 5594.16 & 12.09\% &\textcolor{blue}{32.73\%} & 29.61\% &\textcolor{red}{49.31\%} & 52.46\% & \textcolor{red}{108.48\%} & 116.60\%\\
& sin & \textcolor{blue}{8107.58} & 8027.72 & -0.95\% & \textcolor{red}{45.27\%} & 51.40\% & \textcolor{red}{60.83\%} & 74.06\% & \textcolor{red}{112.50\%} & 117.01\% \\ \hline
\multicolumn{4}{||c|}{\textbf{Average}}   & 7.98\% & 48.68\% & 50.35\% & 63.43\% & 69.46\% &108.57\% & 113.16\% \\ \hline

  \end{tabular}
  \vspace{-0.1in}
  \label{table:area}
\end{table*}

In the second set of experiments, for the remaining six benchmarks in BACS in Table~\ref{table:information}, we generate three RUCA designs with 2 levels, 3 levels and 4 levels respectively. We use 0.1\%, 1\% and 2\% as error thresholds.
In order to highlight the benefits of our methodology, we report  {\it relative power} and {\it total area}. Relative power is defined as the ratio between power of RUCA design (under certain accuracy level) and power of the original circuit. Figure~\ref{fig:result_power} illustrates relative power of RUCAs for each benchmark. Compared to original circuit, RUCA substantially saves power under approximate mode, and use slightly extra power to enable corrector circuit for full-accuracy mode. 
However, as the number of accuracy levels increases, RUCA approximates a given circuit into more configurable blocks, which potentially reduces opportunities to optimize logic synthesis and increases power consumption.


In Table~\ref{table:area}, we thoroughly evaluate all benchmarks in Table~\ref{table:information} and compare the performance against another runtime configurable framework named Approximation through Logic Isolation~\cite{jain2016approximation} in terms of {\it total chip area} and {\it relative power} under each accuracy level including full-accuracy mode. 
We use 3-level runtime configurable designs and set error thresholds as 1\% MAE and 2\% MAE. Red numbers represent that RUCA saves more area or power compared to Logic Isolation. Blue numbers represent that RUCA consumes more area or power.
On average, we are able to save 36.57\% power with 1\% error threshold, and 51.32\% power with 2\% error threshold. To run in full accuracy mode, RUCA consumes 8.57\% more power than the original circuit. However, it is expected  that with approximate computing, the circuits will run approximately most of the time, and only in a few occasions, full accuracy will be needed and enabled.
Compared to Logic Isolation, our RUCA framework has smaller total area in 6 designs out of 9 benchmarks, which on average saves 7.98\% area compared to Logic Isolation. In terms of power consumption, our approach has 4 better results under 2\% error level, and 7 better results under 1\% error level and full-accuracy level respectively. In general, compared to Logic Isolation, RUCA is able to use smaller chip area and consumes less power to implement the same functionality of runtime configurable design, especially under higher-accuracy level. To restore full accuracy, the design overhead of RUCA is relatively less. 



\section{Conclusion}
\label{sec:conclusion}
In this paper, we proposed a novel methodology RUCA to design runtime configurable approximate circuit with Boolean matrix factorization. Factorized matrices are separated to synthesize each approximation block, while a corrector unit is created to restore full accuracy. Moreover, we integrated our methodology with a circuit partitioning and design space exploration scheme to scale our approach, where the algorithm navigates the search space of approximate subcircuits. We evaluated RUCA on a set of benchmarks, and demonstrated that the proposed design significantly saves area and power, while providing flexibility to balance the trade-off between QoR and power. By comparing against Approximation through Logic Isolation, we highlight the state-of-the-art performance of our RUCA approach. In future work, we plan to analyze the influence of different partitioning schemes on RUCA, and improve our methodology with an optimal partitioning strategy.
\newline

\noindent\textbf{Acknowledgments:} {\small This work is partially supported by NSF grant 1814920 and DoD ARO grant W911NF-19-1-0484.}



\bibliographystyle{IEEEtran}
\bibliography{ref}

\begin{thebibliography}{10}
\providecommand{\url}[1]{#1}
\csname url@samestyle\endcsname
\providecommand{\newblock}{\relax}
\providecommand{\bibinfo}[2]{#2}
\providecommand{\BIBentrySTDinterwordspacing}{\spaceskip=0pt\relax}
\providecommand{\BIBentryALTinterwordstretchfactor}{4}
\providecommand{\BIBentryALTinterwordspacing}{\spaceskip=\fontdimen2\font plus
\BIBentryALTinterwordstretchfactor\fontdimen3\font minus
  \fontdimen4\font\relax}
\providecommand{\BIBforeignlanguage}[2]{{%
\expandafter\ifx\csname l@#1\endcsname\relax
\typeout{** WARNING: IEEEtran.bst: No hyphenation pattern has been}%
\typeout{** loaded for the language `#1'. Using the pattern for}%
\typeout{** the default language instead.}%
\else
\language=\csname l@#1\endcsname
\fi
#2}}
\providecommand{\BIBdecl}{\relax}
\BIBdecl

\bibitem{han2013approximate}
J.~Han and M.~Orshansky, ``Approximate computing: An emerging paradigm for
  energy-efficient design,'' in \emph{18th IEEE European Test Symposium}.\hskip
  1em plus 0.5em minus 0.4em\relax IEEE, 2013, pp. 1--6.

\bibitem{salsa}
S.~Venkataramani, A.~Sabne, V.~Kozhikkottu, K.~Roy, and A.~Raghunathan,
  ``Salsa: Systematic logic synthesis of approximate circuits,'' in
  \emph{Design Automation Conference}, 2012, pp. 796--801.

\bibitem{aslan14}
A.~Ranjan, A.~Raha, S.~Venkataramani, K.~Roy, and A.~Raghunathan, ``Aslan:
  Synthesis of approximate sequential circuits,'' in \emph{Design, Automation
  {\&} Test in Europe Conference}, 2014, pp. 1--6.

\bibitem{blasys}
J.~Ma, S.~Hashemi, and S.~Reda, ``{Approximate Logic Synthesis Using Boolean
  Matrix Factorization},'' in \emph{IEEE Transactions on Computer-Aided Design
  of Integrated Circuits and Systems}, 2021.

\bibitem{Nepal14}
K.~Nepal, Y.~Li, R.~I. Bahar, and S.~Reda, ``{ABACUS: A Technique for Automated
  Behavioral Synthesis of Approximate Computing Circuits},'' in \emph{{Design,
  Automation and Test in Europe}}, 2014, pp. 1--6.

\bibitem{Vasicek2016}
Z.~Vasicek and L.~Sekanina, ``Evolutionary design of complex approximate
  combinational circuits,'' \emph{Genetic Programming and Evolvable Machines},
  vol.~17, no.~2, pp. 169--192, 2016.

\bibitem{zervakis2020design}
G.~Zervakis, H.~Amrouch, and J.~Henkel, ``Design automation of approximate
  circuits with runtime reconfigurable accuracy,'' \emph{IEEE Access}, vol.~8,
  pp. 53\,522--53\,538, 2020.

\bibitem{venkataramani2013substitute}
S.~Venkataramani, K.~Roy, and A.~Raghunathan, ``Substitute-and-simplify: A
  unified design paradigm for approximate and quality configurable circuits,''
  in \emph{Design, Automation \& Test in Europe Conference \& Exhibition
  (DATE)}.\hskip 1em plus 0.5em minus 0.4em\relax IEEE, 2013, pp. 1367--1372.

\bibitem{jain2016approximation}
S.~Jain, S.~Venkataramani, and A.~Raghunathan, ``Approximation through logic
  isolation for the design of quality configurable circuits,'' in \emph{Design,
  Automation \& Test in Europe Conference \& Exhibition (DATE)}.\hskip 1em plus
  0.5em minus 0.4em\relax IEEE, 2016, pp. 612--617.

\bibitem{alan2020runtime}
T.~Alan, A.~Gerstlauer, and J.~Henkel, ``Runtime accuracy-configurable
  approximate hardware synthesis using logic gating and relaxation,'' in
  \emph{Design, Automation \& Test in Europe Conference \& Exhibition
  (DATE)}.\hskip 1em plus 0.5em minus 0.4em\relax IEEE, 2020, pp. 1578--1581.

\bibitem{moons2015dvas}
B.~Moons and M.~Verhelst, ``Dvas: Dynamic voltage accuracy scaling for
  increased energy-efficiency in approximate computing,'' in \emph{IEEE/ACM
  International Symposium on Low Power Electronics and Design (ISLPED)}.\hskip
  1em plus 0.5em minus 0.4em\relax IEEE, 2015, pp. 237--242.

\bibitem{BACS}
I.~Scarabottolo, G.~Ansaloni, G.~A. Constantinides, L.~Pozzi, and S.~Reda,
  ``{BACS}: Benchmarks for approximate circuit synthesis,''
  \url{https://github.com/scale-lab/BACS}, 2020.

\bibitem{amaru2015epfl}
L.~Amar{\'u}, P.-E. Gaillardon, and G.~De~Micheli, ``The epfl combinational
  benchmark suite,'' in \emph{24th International Workshop on Logic \&
  Synthesis}, 2015.

\bibitem{reda2019approximate}
S.~Reda and M.~Shafique, \emph{Approximate Circuits}.\hskip 1em plus 0.5em
  minus 0.4em\relax Springer, 2019.

\bibitem{kahng2012accuracy}
A.~B. Kahng and S.~Kang, ``Accuracy-configurable adder for approximate
  arithmetic designs,'' in \emph{Proceedings of the 49th Annual Design
  Automation Conference}, 2012, pp. 820--825.

\bibitem{albicocco2013truncated}
P.~Albicocco, G.~C. Cardarilli, A.~Nannarelli, M.~Petricca, and M.~Re,
  ``Truncated multipliers through power-gating for degrading precision
  arithmetic,'' in \emph{Asilomar Conference on Signals, Systems and
  Computers}.\hskip 1em plus 0.5em minus 0.4em\relax IEEE, 2013, pp.
  2172--2176.

\bibitem{mrazek2018design}
V.~Mrazek, Z.~Vasicek, and L.~Sekanina, ``Design of quality-configurable
  approximate multipliers suitable for dynamic environment,'' in \emph{2018
  NASA/ESA Conference on Adaptive Hardware and Systems (AHS)}.\hskip 1em plus
  0.5em minus 0.4em\relax IEEE, 2018, pp. 264--271.

\bibitem{Miettinen11}
P.~Miettinen and J.~Vreeken, ``Model order selection for boolean matrix
  factorization,'' in \emph{17th ACM SIGKDD international conference on
  Knowledge discovery and data mining}, 2011, pp. 51--59.

\bibitem{miettinen2008discrete}
P.~Miettinen, T.~Mielik{\"a}inen, A.~Gionis, G.~Das, and H.~Mannila, ``The
  discrete basis problem,'' \emph{IEEE transactions on knowledge and data
  engineering}, vol.~20, no.~10, pp. 1348--1362, 2008.

\bibitem{shhmss2016alenex}
S.~Schlag, V.~Henne, T.~Heuer, H.~Meyerhenke, P.~Sanders, and C.~Schulz,
  ``k-way hypergraph partitioning via \emph{n}-level recursive bisection,'' in
  \emph{18th Workshop on Algorithm Engineering and Experiments}, 2016, pp.
  53--67.

\end{thebibliography}

\end{document}